# The sixth Hilbert's problem and the principles of quantum informatics

## Yu. I. Bogdanov

*Institute of Physics and Technology, Russian Academy of Sciences*[*]

By the example of a Fourier transform, the possibilities of Hilbert space geometry applications for statistical model construction are analyzed. In accordance with Bohr's complementarity principle, mutually-complementary coordinate and momentum representations are presented. It was demonstrated that the characteristic function of coordinate distribution may be considered as a convolution of the psi-function in momentum representation and vice versa. The naturalness of coordinate and momentum operators' introduction is demonstrated. A probabilistic interpretation of Hilbert space geometry is given. Cauchy-Bunyakowsky (Cauchy-Schwartz), Cramer-Rao and uncertainty inequalities are considered in the same framework. The principal postulates of quantum informatics as a natural science are presented. It is demonstrated that quantum informatics serves as a theoretic basis for both probability theory and quantum mechanics.

### Introduction

*"Trace everything back to the beginning and you will understand a lot!" (Koz'ma Prutkov «Thoughts and Aphorisms», №247).*

Quantum informatics – a new fundamental field in science - has been actively developed recently. Numerous researches are made on the applications of quantum systems to problems of computation and communication [1-3].

This new field emerged not only because of the practical needs it could provide but it was also due to the internal logic of development of the science. This work attempts to support this thesis by providing the reasoning for the idea that quantum informatics is a natural scientific solution to the so-called sixth Hilbert's problem.

Historically, quantum informatics appeared as a symbiosis of notions of quantum physics (that resulted when studies on atoms, radiation, solid bodies etc. were conducted) and probability theory (that has been widely used in molecular-kinetic theory). However, it would be misleading to believe that quantum informatics simply uses the knowledge accumulated by physics and mathematics. Instead, the theory became the answer to serious conceptual difficulties that physics and statistics encountered.

Let us describe the conceptual difficulties. Everyone who tries to understand the principles of quantum theory using all the available literature, inevitably experiences some serious methodological difficulties. In fact, when quantum theory is usually explained (see [4,5], for example), so-called "peculiarities" in microsystem behaviour are highlighted. One example of such peculiarities is the dual nature of an electron, when it behaves as a particle "on Mondays, Wednesdays and Fridays" and as a wave "on Tuesdays, Thursdays and Saturdays" (while on "Sundays it has a break"). Another example is the Heisenberg microscope, where the measurement of a coordinate of a particle affects its momentum. Wittingly, the uncertainty relation follows from here, just as if the particle could have had both momentum and coordinate before the measurement. Also, the impact of an observer on a physical system and the mysterious transformation rules of classical Poisson brackets into quantum ones are some examples of the peculiarities. Of course, there are many others.

Not surprisingly, having noted such intrinsically contradictory conceptions R. Feynman concluded that quantum mechanics can not be understood in principle. This conclusion is commonly shared by physicists.

To overcome the methodological problems in this work we make the following propositions.

Quantum mechanics indeed can not be understood if it is considered separately from the principles of quantum informatics. In fact, the principles of quantum informatics are simple and natural. We believe that a description of the principles of quantum informatics should precede the description of any physical problems. Quantum physics is merely a physical realization of quantum informatics and not the other way round. In other words, no preliminary study of physical problems can help a person understand the basics of quantum informatics. Conversely, having understood the basics of quantum informatics we may easily comprehend the substance of particular physical problems.

Note that in present work we consider only quantum informatics and do not mention quantum physics. Furthermore, the physical terminology used in this paper is only due to tradition and history. For instance, the uncertainty relation between coordinate and momentum is a mathematical rather than physical phenomenon. The notion of momentum corresponds to the variable that emerges when a Fourier transform upon a state vector is made (i.e. the momentum is not a mere product of mass and velocity). Similarly, the fact that the uncertainty relation holds the name of Heisenberg is only a matter of appreciation to one of the founders of quantum theory, but not an attempt to present the thought experiments by Heisenberg as a copy-book truth.

---

[*] E- mail: bogdan@ftian.oivta.ru

The conceptual physical difficulties that quantum physics encountered during its development are well known in general. At the same time, similar difficulties that classical statistics experiences when it attempts to describe physical systems are less carefully explored. In present work, an attempt has been made to show that the principles of quantum informatics have statistical nature. Note, however, that methodologically, the probabilistic nature of quantum theory is fundamentally different from the statistical nature of classical probability theory. That is due to the fact that in classical theory the notion of probability is subjective (it is natural that the notion of classical probability arose from the analysis of games of chances). It is well-known that all problems of classical statistics can be reduced to the classical scheme containing a basket (urn) with $N$ balls, where $M$ balls are white, and to its more abstract variants. It is claimed that the probability of selecting a white ball is equal to $P = M/N$.

Such a comprehension of probability is definitely subjective. The process looks like a random one only to a person that has his eyes blindfolded. The nature of the phenomenon is not random at all. Another person that watches the process using a video-camera inside the basket knows definitely which ball is to be taken. It is reasonable that the Nature does not care what an observer knows and what he does not know. Therefore, such theories can not describe objective physical reality. That is the source of the problems of classical probability theory and statistical physics (the latter, in fact, simply brings pseudo-randomness into classical mechanics). In scientific framework, probability theory should describe mathematical models of events that are random by their nature, rather than because of different subjective opportunities of individuals. Such models are considered in quantum mechanics. Let us briefly note, without going into details that will be provided below, that the mathematical model of quantum informatics implies a proper division into a preparation of a quantum state and its measurement.

The vector of a quantum state is an objective probabilistic characteristic of a quantum system. From the one hand, the vector accumulates all the physical parameters that are related to a quantum state preparation. From the other hand, a single state vector corresponds to a variety of measurement schemes. As a result, it appears that a quantum state can not be reduced to any subjective classical sampling scheme and, therefore, it describes the objective randomness in the Nature.

The work is organized as follows. First of all, a brief introduction to the sixth Hilbert's problem is given. Then, a probabilistic interpretation of Hilbert space vectors is provided (by the example of mutually-complementary Fourier transforms). The considered approach based on a combination of statistical principles and Bohr's complementarity principle has been described by the author in previous papers [6-10]. The approach allows one to shed light on some issues of classical mathematical statistics in a different framework. In particular, it shows that the notions of momentum and coordinate operators are in some way embedded in classical statistics. However, in a quantum informatics framework, the insufficiency of the classical statistical approach is that it considers only one of the possible mutually-complementary distributions, ignoring the others. As a result, the essential notion of a state vector can not be derived in classical statistical theory, which makes it rather limited and unattractive.

In the second part of the work the issues of the precision of Hilbert space statistical characteristics are considered.

In the third part of the work the postulates of quantum informatics are formulated. Here our choice of postulates is analogous to the choice in the well-known book by Nielsen and Chuang [1], however we also stress the statistical nature of every postulate.

In the discussion part the history of the sixth Hilbert's problem evolution in the XX century is briefly discussed.

Finally, in the conclusion the main findings of the work are summarized.

## The motivation. What is the 6$^{th}$ Hilbert's problem?

In the famous report by D. Hilbert on the 8$^{th}$ of August 1900 in Paris at the Second International Congress of Mathematicians, a number of problems were formulated that had a significant impact on the development of mathematics and related sciences in the XX century.

Hilbert formulated as many as 23 problems, and of most interest for us is the 6$^{th}$ problem, which was formulated as a "mathematical description of axioms of physics". "The research of the basis of geometry are closely related to axiomatic construction of the physical sciences, where mathematics plays a vital role, i.e. probability theory and mechanics.

Concerning probability theory axioms, I would like the logical explanation of the theory to be developed in parallel with the method of mean values in mathematical physics and the kinetic gas theory, in particular. (a translation from the Russian edition [11] p.415).

Today, after more that one hundred years have passed, we may say that Hilbert was certainly right.

Note that the mathematical formalization of the basics of probability theory are connected with the micro-object science, which was the molecular-kinetic theory at that times. Astonishingly, just a few months after the report by Hilbert, Max Planck made a report that was the origin of the new quantum era on the 14$^{th}$ of December 1900.

While Hilbert states that the axiomatic theory has to be constructed by analogy with geometry, we will see that the Hilbert space geometry is the basis of quantum informatics. In particular, the scalar product of vectors in Hilbert space $|\langle \psi_{out} | \psi_{in} \rangle|^2$ may be interpreted as the probability of *detecting* a quantum system in state $\psi_{out}(x)$ if it was originally *prepared* in state $\psi_{in}(x)$. Also, Hilbert predicted that The Lie's theory of group transformations may play an important role in constructing the unifying theory ([11], p.416), and, actually, the importance of Lie's theory in quantum informatics is well known.

Finally, note that Hilbert is concerned by a mathematically strict transition from micro-world to the macro-world. There, he relates to the book by Boltzmann on the principles of classical mechanics. Still, even now little is done on this issue. ([12]).

The actuality of the sixth Hilbert's problem was defined by the state of science between XIX and XX. The famous H-theorem formulated by Boltzmann in 1872 [13], was heavily argued by many scholars, for example Poincare. The difficulty was due to the contradiction between the reversibility of classical mechanics' laws and the non-reversibility of the second law of thermodynamics. Though Boltzmann's comments on this issue were rather convincing, still any symbiosis of classical mechanics and statistics was internally contradictory in mathematical framework. Note, however, that Boltzmann's approach to statistical thermodynamics was not purely classical because he used the notion of energy quantization for the sake of methodology (28 years before Planck) [13]. As a result, any attempts to combine mechanics and statistics should have led to quantum notions. Therefore, it was clear for Hilbert and many other scholars that mechanics, probability theory and thermodynamics could not have evolved independently. By formulating his sixth problem Hilbert attempted to overcome the difficulty by means of axioms and obtain a general non-contradictory theory. As we shall see quantum informatics may be that theory.

## 1. Fourier transform. Analysis of mutually-complementary statistical variables.

The nature of quantum informatics can be demonstrated by the example of statistical properties of complex functions and their Fourier transform.

## 1.1. Statistical interpretation of direct and inverse Fourier transform. Coordinate and momentum distribution.

Let $\psi(x)$ be an arbitrary complex function defined in Hilbert space $L_2$. It has a finite integral of the square of the absolute value.

$$\int |\psi(x)|^2 dx < \infty$$

Direct and inverse Fourier transforms are given by:

$$\psi(x) = \frac{1}{\sqrt{2\pi}} \int \widetilde{\psi}(p) \exp(ipx) dp \tag{1.1}$$

$$\widetilde{\psi}(p) = \frac{1}{\sqrt{2\pi}} \int \psi(x) \exp(-ipx) dx \tag{1.2}$$

It is well-known that for a function and its Fourier transform the following Parseval equation is satisfied.

$$\int |\psi(x)|^2 dx = \int |\widetilde{\psi}(p)|^2 dp$$

In the statistical framework, the Parseval equation defines the condition of normalization conservation while transiting from coordinate state vector representation to momentum representation.
The distribution density in the prior (coordinate) representation is:

$$P(x) = |\psi(x)|^2 \tag{1.3}$$

The Fourier transform $\widetilde{\psi}(p)$ defines momentum probability distribution

$$\widetilde{P}(p) = |\widetilde{\psi}(p)|^2 \tag{1.4}$$

The Parseval equation states that total probability is independent from the choice of representation. The total probability can be normalized to any positive value that is usually unity.

$$\int P(x) dx = \int \widetilde{P}(p) dp = 1 \tag{1.5}$$

The considered equation implies that the total probability is equal to unity. Note that in research other normalization conditions are also used. In particular, in decay problem and micro-objects scattering, the total probability can be characterized by the total number of events per one unit of time.

Note that function $\psi(x)$ and its Fourier transform $\tilde{\psi}(p)$ contain equal information (the knowledge of one of them allows one to derive the other one). They are called probability amplitudes in coordinate and momentum representations correspondingly (sometimes other terms are used – wave function, psi-function, state vector).

### 1.2 Bohr's complementarity principle.

Coordinate $P(x)$ and momentum $\tilde{P}(p)$ distributions are called mutually-complementary statistical distributions, because they statistically complement one another. During the measurement, the information about the phase of the wave function $\psi(x)$ is lost. Actually, during the transition from $\psi(x)$ to $\psi(x)\exp(iS(x))$, where $S(x)$ is an arbitrary real function (phase), the coordinate probability distribution $P(x)$ is not affected. However, such a transition may have an impact on momentum representation $\tilde{P}(p)$. In this sense, $\tilde{P}(p)$ comprises additional information compared to $P(x)$.

Bohr's complementarity principle had much effect on the considered terminology. According to quantum theory, the information of a statistical quantum system is embedded in the wave function (state vector) $\psi(x)$. At the same time, for an experimental extracting of the information, it is not sufficient to use only one fixed representation. For the description of a quantum system to be complete, one has to conduct measurements that satisfy a set of mutually-complementary distributions. That is the statistical idea of Bohr's complementarity principle. Coordinate and momentum distributions are an example of such mutually-complementary distributions. The statistical complementarity principle plays a vital role in problems of quantum informatics.

The model of quantum informatics assumes that there are some fixed "rules of the game" between the Nature and the man. A psi-function (generally, complex) comprises full information about a quantum system. However, due to the statistical nature of quantum mechanics, we do not have the possibility to measure it directly. The only thing that we can do is to conduct upon a given number of representatives in the same state $\psi(x)$.

Experimentally, examining the validity of quantum theory is based on reconstructing (by statistical measurements) the properties of a state vector in Hilbert space that are unavailable for direct observation. So far, all experiments that have been conducted were in accordance with quantum informatics.

Note that traditional probability theory and mathematical statistics describe only single (not mutually-complementary) probability distributions (one- and multi-dimensional). The principle of complementarity violates the so-called composite random variables axiom [14]. According to the axiom, there is only one way of transition from singular properties description to the description of a set of such properties. This approach is based on the transition from one-dimensional distributions to multi-dimensional distributions. In quantum informatics this is not the case. For mutually-complementary distributions, their set is not a distribution but a more general object, i.e. quantum state. For instance, no composite distribution $P(x,p)$ corresponds to coordinate $P(x)$ and momentum $\tilde{P}(p)$ distributions. Such a distribution would violate the Heisenberg's uncertainty principle, as we will see below. According to quantum informatics, a state vector in Hilbert space unifies all mutually-complementary distributions.

### 1.3. Characteristic function. Mean value and moments' calculation. Incompleteness of classical statistics and completeness of quantum statistics.

Coordinate and momentum representations of a state vector are equivalent. Therefore, it is quite interesting to consider the properties of coordinate probability distribution in momentum representation framework and the other way round.

Coordinate probability distribution in momentum representation can be expressed as follows:

$$P(x) = \psi^*(x)\psi(x) = \frac{1}{2\pi} \int dp\, dp_1 \tilde{\psi}^*(p)\tilde{\psi}(p_1)\exp(-ix(p-p_1)) =$$

$$\frac{1}{2\pi} \int du\, dp\, \tilde{\psi}^*(p)\tilde{\psi}(p-u)\exp(-ixu) = \frac{1}{2\pi} \int f(u)\exp(-ixu)du \tag{1.6}$$

In the latter equation the characteristic function was introduced:

$$f(u) = \int dp\, \tilde{\psi}^*(p)\tilde{\psi}(p-u) = \int dp\, \tilde{\psi}^*(p+u)\tilde{\psi}(p) \tag{1.7}$$

As a result, distribution density can be considered as Fourier inverse transformation of the characteristic function

$$P(x) = \frac{1}{2\pi} \int f(u)\exp(-ixu)du \tag{1.8}$$

Then, the characteristic function is a direct Fourier transform of density, or, equally, mathematical expectance (mean value) of random value $\exp(iux)$:

$$f(u) = \int P(x)\exp(ixu)dx = M(\exp(iux)) \tag{1.9}$$

Similarly, it can be shown that the characteristic function can be expressed using convolution of the coordinate psi-function. Let $\tilde{f}(t)$ be the characteristic function of momentum distribution that is a Fourier-transform of momentum distribution density $\tilde{P}(p)$:

$$\tilde{f}(t) = \int \tilde{P}(p)\exp(ipt)dp = M(\exp(ipt)) \tag{1.10}$$

In analogy to (1.7) the characteristic function under consideration can be expressed as a convolution of coordinate psi-function.

$$\tilde{f}(t) = \int dx\, \psi^*(x-t)\psi(x) = \int dx\, \psi^*(x)\psi(x+t) \tag{1.11}$$

However, not every function can be considered as a characteristic function, because inverse Fourier transform of the characteristic function has to yield a real non-negative function (unity-normalized density).

The calculations made above can explain the following statement: for the function $f(u)$ to be a characteristic function it is necessary and sufficient it to be represented as a convolution of a complex function $\tilde{\psi}(p)$ that satisfies the normalization condition:

$$\int dp\, |\tilde{\psi}(p)|^2 = 1 \tag{1.12}$$

Necessity: Let $f(u)$ be a characteristic function. Then, according to (1.8) it defines some density $P(x)$. Let us define a psi-function as $\psi(x) = \sqrt{P(x)}\exp(iS(x))$, where $S(x)$ is an arbitrary real function (for instance, $S(x) = 0$). The considered procedure may be called an completion of a classical statistic distribution to a quantum state. The function $\tilde{\psi}(p)$, defined by the inverse Fourier transform (1.2) provides the desired decomposition of the characteristic function as a convolution (1.7). Thus, one can put into correspondence to any characteristic function a wave function in momentum space. Furthermore, such a representation is ambiguous

Sufficiency: Let $f(u)$ be represented as a convolution (1.7) of some function $\tilde{\psi}(p)$, normalized according to (1.1), while the distribution density $P(x)$ is normalized as (1.3). Then, $f(u)$ will be a characteristic function of distribution $P(x)$. Therefore, one may establish a correspondence between any wave function of momentum space $\tilde{\psi}(p)$, a characteristic function $f(u)$ and a unique distribution $P(x)$. Proof completed.

Even in classical statistics the equation (1.7) implicitly reveals the existence of momentum space and the corresponding wave function $\widetilde{\psi}(p)$. This relation characterizes the scarcity of classical notions of probability. Actually, equation (1.7) can not derive a unique wave function $\widetilde{\psi}(p)$ in momentum space. Similarly, relation (1.3) can not derive $\psi(x)$ unambiguously in coordinate space. Therefore, one classical probability distribution may be described by a number of different quantum objects, while one can avoid this obstacle introducing Hilbert space state vectors.

As already noted above, for the statistical theory to be complete it has to be expanded in a way so that a probability distribution transforms into a quantum state vector (for example, a phase multiplier may be introduced).

At the same time, quantum objects of a more profound nature completely fit to the structure of a state vector Hilbert space. A quantum state vector does not need (and even does not allow for) any complements to any objects of a more general nature.

Postulates of quantum description will be described in section 3, while here let us note the following fundamental difference between a probability distribution description and a state vector description.

In the framework of classical probability theory, let us assume that variables $x_1, x_2, ..., x_s$ are interconnected by some probability distribution $P(x_1, x_2, ..., x_s)$. The existence of such distribution does not exclude the existence of additional $r$ variables $x_{s+1}, x_{s+2}, ..., x_{s+r}$, so that the former variables are statistical dependent on these variables. The two groups of variables are statistical dependent if a joint distribution of dimension $s + r$ is non-separable (can not be factorized), i.e. it can not be represented as a product of two distributions of dimensions $s$ and $r$, i.e.

$$P(x_1, x_2, ..., x_s, x_{s+1}, ..., x_{s+r}) \neq P(x_1, x_2, ..., x_s) P(x_{s+1}, x_{s+2}, ..., x_{s+r}).$$

The idea behind this relation is that any dependence discovered between the initial variables $x_1, x_2, ..., x_s$ may indeed be a fiction, as real physical reasons could be determined by additional ("hidden") variables $x_{s+1}, x_{s+2}, ..., x_{s+r}$. Thus, generally speaking, any classical statistical analysis can not make objective scientific deductions. As far as 100 years ago, Bernard Show derided the situation by writing that statisticians could easily have proven that wearing top hats would enlarge life and protect from illnesses. This flaw in classical statistics is well known [15] and conscientious researchers consider statistical analysis as an ancillary tool only.

It is worth noting that quantum theory does not have a similar flaw. Let variables $x_1, x_2, ..., x_s$ make a quantum state $\psi(x_1, x_2, ..., x_s)$. Then it is impossible for the variables to be statistically dependent on any other variables in the Universe including those "hidden" variables inside the system.

In other words, expanding the initial system by including any additional variables $x_{s+1}, x_{s+2}, ..., x_{s+r}$ will necessarily produce a separable composite quantum state, i.e. it will always be possible to represent a quantum state as a product of independent state vectors:

$$\psi(x_1, x_2, ..., x_s, x_{s+1}, ..., x_{s+r}) = \psi(x_1, x_2, ..., x_s) \psi(x_{s+1}, x_{s+2}, ..., x_{s+r}).$$

For instance, if one is to introduce a spin to non-relativistic quantum mechanics then the state vector becomes the product of coordinate and spin functions. Of course, such a factorization of a quantum state that results in independent internal and external variables is a kind of simplification. Still, such simplifications make the basis of any science.

Now, suppose that the considered state is non-separable. i.e.

$$\psi(x_1, x_2, ..., x_s, x_{s+1}, ..., x_{s+r}) \neq \psi(x_1, x_2, ..., x_s) \psi(x_{s+1}, x_{s+2}, ..., x_{s+r}).$$

Then it is impossible to assign any state vectors to sub-systems $x_1, x_2, ..., x_s$ and $x_{s+1}, x_{s+2}, ..., x_{s+r}$. Such systems can not be considered as independent closed systems, no matter how far they are from one another. A well-known example of the fact is an EPR-state. Thus, the notion of closure of a physical system in quantum theory significantly differs from that in classical theory.

Spatial isolation can not serve as a feature of closure. Instead, in quantum theory there is an internal statistical criterion – a complete description that is independent from all other variables is only possible for systems that may be described by a state vector. Ironically, EPR states, contrary to their developers' expectations, are an important argument in favour of completeness of quantum theory and definitely not against it.

The arguments presented reveal the incompleteness of the classical probability theory and the completeness of the quantum theory. Note, however, that the incompleteness of classical probability theory is well-known and it is not considered as a disadvantage by the specialists in the theory [16].

We believe, however, that incompleteness of classical probability theory is its drawback. It is formally eliminated by expanding classical probability distribution to a quantum state vector as described above. Also, note that an incomplete description is often used in quantum theory too, where it corresponds to density matrix apparatus. The necessity to introduce a density matrix is due to the fact that a quantum physical system can often interact with its environment in a complicated way. Still, formally any density matrix may be expanded to a pure state (the so-called quantum state purification procedure).

The representation of a characteristic function as a convolution is a well-known result in classical probability theory (see theorem 4.2.4 in [17]). Nonetheless, the classical theory does not describe the nature of complex function $\tilde{\psi}(p)$ that appears in this representation. Also, the applications of the theorem to quantum mechanics are explained in [18]. Nevertheless, the findings of the book may be misleading. For instance, we doubt the statement that "quantum-mechanical probabilistic description is completely embedded in the classical probability theory" ([18], p.20). Such a conclusion ignores Bohr's complementarity principle as was discussed above.

Let us briefly describe the way to calculate moments of a random value by means of the characteristic function. The value of the characteristic function in zero point is equal to unity.

$$f(0) = 1 \tag{1.13}$$

The moments of a random value can be expressed by the corresponding derivatives of the characteristic function at zero point.

$$f'(u) = \int P(x)\exp(ixu)ix\,dx \text{, thus,}$$
$$f'(0) = iM(x). \tag{1.14}$$

As a result, the first derivative of the characteristic function in zero point is related to the mean value.

Similarly, one may derive the fact that the $k$-th derivative is related to the $k$-th moment as:

$$f^{(k)}(0) = i^k M(x^k), \quad k = 0,1,2,... \tag{1.15}$$

## 1.4. Coordinate and momentum operators in coordinate and momentum representations. Fundamental commutation relations.

The properties of the characteristic function allow one to introduce coordinate operator in momentum representation:

$$M(x) = -if'(0) = -i\int dp\,\tilde{\psi}^*(p)\frac{\partial}{\partial u}\tilde{\psi}(p-u)_{u=0} = i\int dp\,\tilde{\psi}^*(p)\frac{\partial}{\partial p}\tilde{\psi}(p) \tag{1.16}$$

Therefore, while in coordinate representation the coordinate is described by operator $\hat{x}$, i.e. $\hat{x}\psi(x) = x\psi(x)$, in momentum representation the coordinate operator takes the form $\hat{x} = i\frac{\partial}{\partial p}$.

Similarly, it may be shown that in momentum representation momentum is described by operator $\hat{p}$, i.e. $\hat{p}\tilde{\psi}(p) = p\tilde{\psi}(p)$, while in coordinate representation by $\hat{p} = -i\frac{\partial}{\partial x}$.

Note that the operators of coordinate and momentum are Hermitian. During a transition from one representation to the other makes the following commutation relation is invariant.

$$\hat{p}\hat{x} - \hat{x}\hat{p} = -i \tag{1.17}$$

For many degrees of freedom the equation takes the form:

$$\hat{p}_j\hat{x}_k - \hat{x}_k\hat{p}_j = -i\delta_{jk}, \quad j,k = 1,2,...,s \tag{1.18}$$

Where $s$ is the number of degrees of freedom.

The considered relation implies that every momentum does not commute with its canonically-conjugate coordinate, but commutes with all other coordinates.

All coordinates and all momentums commute with each other.

$$\hat{x}_j \hat{x}_k - \hat{x}_k \hat{x}_j = 0 \quad (1.19)$$

$$\hat{p}_j \hat{p}_k - \hat{p}_k \hat{p}_j = 0 \quad (1.20)$$

The transforms that do not affect the fundamental commutation relations are called canonical.

Fourier transform is a particular case of unitary transformations. It appears that one may consider arbitrary unitary transformations in quantum informatics. This will change the representation to the complementary representation. Measurements in these representations provide a set of mutually-complementary distributions. These considerations form the basis for postulates of quantum informatics (see chapter 3).

## 2. Precision of statistical characteristics and the Heisenberg's uncertainty principle

Below it will be demonstrated that Cauchy-Bunyakowsky inequalities, uncertainty inequalities and Cramer-Rao inequalities can be derived from the same mathematical approach that implies non-negativity of some quadratic trinomial.

In sections 2.1-2.3 the elementary notions related to the Cauchy-Bunyakowsky and uncertainty inequalities are explained using the main principles of quantum informatics. In part 2.4 a so-called Schrödinger-Robertson uncertainty relation is presented. In section 2.5 multi-dimensional inequality relations are considered. In sections 2.6, 2.7 and 2.6 the notions of Fisher information and Cramer-Rao inequalities that are well known in classical mathematical statistics are studied in the new framework of quantum information.

### 2.1. Cauchy-Bunyakowsky inequality for state vectors and its statistical interpretation.

The inequality relation considered can be studied for arbitrary linear spaces, where the notion of scalar product is introduced. Let us provide some examples of such spaces.

In complex finite dimensional space $C^s$ of dimension $s$ the scalar product of two vectors is defined by the following equation in Dirac notation:

$$\langle \varphi | \psi \rangle = \sum_{j=1}^{s} \varphi_j^* \psi_j \quad (2.1)$$

In infinite dimensional Hilbert space $l_2$ the similar definition takes the form:

$$\langle \varphi | \psi \rangle = \sum_{j=1}^{\infty} \varphi_j^* \psi_j \quad (2.2)$$

Finally, if $\psi(x)$ and $\varphi(x)$ - are complex functions in space $L_2$, their scalar product is:

$$\langle \varphi | \psi \rangle = \int \varphi^*(x) \psi(x) dx \quad (2.3)$$

Let us demonstrate that for all vectors in linear space with scalar products the following relation (Cauchy-Bunyakowsky inequality) is satisfied:

$$|\langle \varphi | \psi \rangle|^2 \leq \langle \varphi | \varphi \rangle \langle \psi | \psi \rangle \quad (2.4)$$

For definiteness let us consider functions in space $L_2$.

Also, let us assume that $\langle \varphi | \psi \rangle$ is a real number.

Let $\xi$ be a real parameter. Let us consider the following wittingly non-negative function $\xi$ (this function is the integral of a wittingly non-negative expression)

$$F(\xi) = \int (\psi(x) + \xi \varphi(x))(\psi^*(x) + \xi \varphi^*(x)) dx \geq 0 \quad (2.5)$$

In Dirac notation:

$$F(\xi) = (\langle \psi | + \xi \langle \varphi |)(|\psi\rangle + \xi |\varphi\rangle)$$

In expanded notation the considered function is a quadratic trinomial:

$$F(\xi) = \xi^2 \langle \varphi | \varphi \rangle + 2\xi \langle \varphi | \psi \rangle + \langle \psi | \psi \rangle \tag{2.6}$$

Here we have used the assumption of reality of scalar product. i.e. $\langle \varphi | \psi \rangle = \langle \psi | \varphi \rangle$

The condition for non-negativity implies that the discriminant of the expression is less of equal to zero.

$$4(\langle \varphi | \psi \rangle)^2 - 4\langle \varphi | \varphi \rangle \langle \psi | \psi \rangle \leq 0 \tag{2.7}$$

Thus in this case the Cauchy-Bunyakowsky inequality is satisfied:

$$(\langle \varphi | \psi \rangle)^2 \leq \langle \varphi | \varphi \rangle \langle \psi | \psi \rangle \tag{2.8}$$

Let us assume that $\langle \varphi | \psi \rangle$ is a complex number.

Let $\langle \varphi | \psi \rangle = r \exp(i\alpha)$, where $r$ and $\alpha$ are real numbers.

Then, let us introduce a function that differs from $\varphi(x)$ only by the phase multiplier.

$$\tilde{\varphi}(x) = \varphi(x) \exp(i\alpha)$$

Then $\langle \tilde{\varphi} | \psi \rangle = r$ is a real number and the following inequality proved above is satisfied:

$$(\langle \tilde{\varphi} | \psi \rangle)^2 \leq \langle \tilde{\varphi} | \tilde{\varphi} \rangle \langle \psi | \psi \rangle$$

Accounting for the fact that the introduced phase transformation does not alter the scalar product absolute value, we get the following relation:

$$|\langle \varphi | \psi \rangle|^2 = (\langle \tilde{\varphi} | \psi \rangle)^2, \quad \langle \tilde{\varphi} | \tilde{\varphi} \rangle = \langle \varphi | \varphi \rangle$$

Therefore, Cauchy-Bunyakowsky inequality is satisfied in the general case also:

$$|\langle \varphi | \psi \rangle|^2 \leq \langle \varphi | \varphi \rangle \langle \psi | \psi \rangle \tag{2.9}$$

Let us introduce value $F$, called the fidelity of quantum states $\varphi(x)$ and $\psi(x)$.

$$F = \frac{|\langle \varphi | \psi \rangle|^2}{\langle \varphi | \varphi \rangle \langle \psi | \psi \rangle} \tag{2.10}$$

For normalized state vectors we have:

$$F = |\langle \varphi | \psi \rangle|^2 \tag{2.11}$$

The Cauchy-Bunyakowsky inequality implies that:

$$0 \leq F \leq 1 \tag{2.12}$$

Thus, from this relation it is natural to consider that $F$ defines some kind of probability. That is really the case. Statistically, $F$ defines the probability that a quantum system, prepared in state $\psi(x)$, will be detected in state $\varphi(x)$.

In nature, information transmission implies that the state $\psi(x)$, prepared on the one side of transmission channel (transmitter system) can be detected in the same state on the other side of the channel (receiver system). In the ideal case the receiver can be set for detecting the same quantum state when $\varphi(x) = \psi(x)$ (accurate to the phase multiplier). Then $F = 1$. In reality, the sender system state $\psi(x)$ and the receiver system state $\varphi(x)$ are at least a bit different and $F < 1$. Thus, in this case $F$ defines the probability of success of state transmission.

## 2.2 Cauchy-Bunyakowsky inequality for random values

Let $Y = Y(x)$ and $Z = Z(x)$ be real random values defined by some arbitrary functions of coordinate $x$. Let $\xi$ be a real parameter. Consider some wittingly non-negative function of $\xi$:

$$F(\xi) = \langle \psi | (\xi Y + Z)^2 | \psi \rangle \geq 0 \tag{2.13}$$

In expanded notation the considered function is a quadratic trinomial:

$$F(\xi) = \xi^2 M(Y^2) + 2\xi M(YZ) + M(Z^2) \tag{2.14}$$

The condition for non-negativity implies that the discriminant is less or equal to zero:

$$4(M(YZ))^2 - 4M(Y^2)M(Z^2) \leq 0 \tag{2.15}$$

Therefore, for all commuting random values the Cauchy-Bunyakowsky inequality is satisfied:

$$(M(YZ))^2 \leq M(Y^2)M(Z^2) \tag{2.16}$$

In particular, if we consider $Y - M(Y)$ and $Z - M(Z)$ as random values with zero averages, then the variances satisfy the following relation:

$$D_Y D_Z \geq \left[ M\big((Y - M(Y))(Z - M(Z))\big) \right]^2 \tag{2.17}$$

Thus the correlation coefficient is

$$r^2 \leq 1 \tag{2.18}$$

In general the correlation coefficient is defined by the following equation:

$$r = \frac{M\big[(Y - M(Y))(Z - M(Z))\big]}{\sqrt{D_Y D_Z}} \tag{2.19}$$

The square of correlation coefficient is sometimes called determination coefficient. This coefficient describes how much random value $Y$ determines random value $Z$ and the other value round. It may be shown that the Cauchy-Bunyakowsky inequality transforms into equality if and only if the random variables $Y$ and $Z$ are linearly dependent.

## 2.3. Heisenberg's uncertainty principle for coordinate and momentum

Let us modify the example presented above. Consider expression $\xi \frac{\partial}{\partial x} + x$ instead of $\xi Y + Z$. Note that the operator of differentiation is not Hermitian, because $\frac{\partial^+}{\partial x} = -\frac{\partial}{\partial x}$. To make this expression more demonstrative let us introduce Hermitian momentum operator $\hat{p} = -i \frac{\partial}{\partial x}$.

Let us consider a wittingly non-negative function of $\xi$ as we did before.

$$F(\xi) = \langle \psi | (-i\xi \hat{p} + \hat{x})(i\xi \hat{p} + \hat{x}) | \psi \rangle \geq 0 \tag{2.20}$$

In expanded notation:

$$F(\xi) = \xi^2 M(\hat{p}^2) - i\xi M(\hat{p}\hat{x} - \hat{x}\hat{p}) + M(\hat{x}^2) \tag{2.21}$$

We also account for the canonical commutation relation

$$\hat{p}\hat{x} - \hat{x}\hat{p} = -i \tag{2.22}$$

Let us consider values $\hat{x} - M(\hat{x})$ and $\hat{p} - M(\hat{p})$ as observable, which, of course, satisfy the same commutation relation. Then we can derive the Heisenberg uncertainty relation for coordinate and momentum variances.

$$D_x D_p \geq \frac{1}{4} \tag{2.23}$$

Momentum variance is the mean square of momentum minus the mean momentum squared:

$$D_p = M(\hat{p}^2) - (M(\hat{p}))^2 \tag{2.24}$$

In expanded notation the mean square of momentum is:

$$M(\hat{p}^2) = -\int \psi^*(x) \frac{\partial^2}{\partial x^2} \psi(x) dx = \int \frac{\partial}{\partial x} \psi^*(x) \frac{\partial}{\partial x} \psi(x) dx \tag{2.25}$$

From that it follows that the inequality transfers into an equality if and only if:

$$(i\xi\hat{p} + \hat{x})|\psi\rangle = 0 \tag{2.26}$$

for some $\xi$.

This equality holds true only for a Gauss-type state (the ground state of a harmonic oscillator).
The solution of this equation in coordinate and momentum representations are:

$$\psi(x) = \frac{1}{(2\pi\sigma_x^2)^{1/4}} \exp\left(-\frac{(x-x_0)^2}{4\sigma_x^2}\right) \tag{2.27}$$

$$\tilde{\psi}(p) = \frac{1}{(2\pi\sigma_p^2)^{1/4}} \exp\left(-\frac{(p-p_0)^2}{4\sigma_p^2}\right) \tag{2.28}$$

Here $x_0$ and $\sigma_x^2$ - are the mean value and the variance for coordinate distribution correspondingly, while $p_0$ and $\sigma_p^2$ are the mean value and the variance for momentum distribution.

The variance of coordinate and momentum of the derived Gauss state are defined by the introduced parameter $\xi$.

$$\sigma_x^2 = \frac{\xi}{2}, \quad \sigma_p^2 = \frac{1}{2\xi} \tag{2.28}$$

Thus, the considered values are interconnected with each other by the minimum uncertainty relation.

$$\sigma_x^2 \sigma_p^2 = \frac{1}{4} \tag{2.30}$$

As a result, the state that minimizes the uncertainty relation is described by a real function. This is not accidental. It is evident that adding an arbitrary phase multiplier to a real psi-function can not decrease the variance of momentum and therefore enforce the inequality considered.

### 2.4 Schrödinger-Robertson uncertainty relation.

The inequality proposed by Schrödinger and Robertson possesses the properties of both the Cauchy-Bunyakowsky inequality and the Heisenberg's uncertainty relation and in some sense it can be considered as a generalization of the two [19,20].

Let $z_1$ and $z_2$ be two arbitrary observables. Without loss of generality we may consider them as centered: $M(z_1) = M(z_2) = 0$.

Let us consider the following wittingly non-negative expression:

$$F(\xi) = \langle \psi | (\xi \exp(-i\varphi)z_2 + z_1)(\xi \exp(i\varphi)z_2 + z_1) | \psi \rangle \tag{2.31}$$

Here $\xi$ is an arbitrary real value, $\varphi$ is a fixed real value (the phase which we will choose later).
Let us define the covariance of values as

$$\text{cov}(z_1, z_2) = \frac{1}{2}\langle\psi|z_1 z_2 + z_2 z_1|\psi\rangle \quad (2.32)$$

Note that we needed to make this expression symmetric for the corresponding operator to be Hermitian.
Let:

$$z_1 z_2 - z_2 z = iC, \quad (2.33)$$

where $C$ is a Hermitian operator. Then:

$$M(C) = -i\langle\psi|z_1 z_2 - z_2 z_1|\psi\rangle \quad (2.34)$$

In expanded notation the expression for $F(\xi)$ has the form:

$$F(\xi) = \xi^2 M(z_2^2) + \xi(2\text{cov}(z_1, z_2)\cos(\varphi) - M(C)\sin(\varphi)) + M(z_1^2) \quad (2.35)$$

Let:

$$\rho^2 = 4(\text{cov}(z_1, z_2))^2 + (M(C))^2, \quad (2.36)$$

It is evident that it is possible to choose such an angle $\beta$, so that the following identity is satisfied:

$$2\text{cov}(z_1, z_2) = \rho\cos(\beta) \quad (2.37)$$
$$M(C) = \rho\sin(\beta) \quad (2.38)$$

Then:

$$F(\xi) = \xi^2 M(z_2^2) + \xi\rho\cos(\varphi + \beta) + M(z_1^2) \geq 0 \quad (2.39)$$

Let us choose the phase $\varphi$, so that $\cos(\varphi + \beta) = 1$. This choice evidently provides the strictest inequality:

$$M(z_1^2)M(z_2^2) = D(z_1)D(z_2) \geq \frac{\rho^2}{4} = \left((\text{cov}(z_1, z_2))^2 + \frac{(M(C))^2}{4}\right) \quad (2.40)$$

Let us define the correlation coefficient between observables $z_1$ and $z_2$ as:

$$r = \frac{\text{cov}(z_1, z_2)}{\sqrt{D(z_1)D(z_2)}} \quad (2.41)$$

As a result, the Schrödinger-Robertson inequality takes the form:

$$D(z_1)D(z_2) \geq \frac{(M(C))^2 K^2}{4}, \quad (2.42)$$

where $K = \frac{1}{\sqrt{1 - r^2}}$ \quad (2.43)

The introduced $K$ is analogous to the well-known Schmidt number [21]. This number is of fundamental importance for quantum correlation and quantum information description.
Now let us considered the momentum and coordinate operators as observables:

$$z_1 = x, \quad z_2 = p.$$

Then, due to the fundamental interchange relation for coordinate and momentum, $C$ is a unitary operator.
In this case the Schrödinger-Robertson uncertainty relation takes the following form:

$$D(x)D(p) \geq \frac{K^2}{4} \quad (2.44)$$

Let $\Delta x = \sqrt{D(x)}$, $\Delta p = \sqrt{D(p)}$ be the uncertainties (standard deviations) for coordinate and momentum. Then:

$$\Delta x \Delta p \geq \frac{K}{2} \tag{2.45}$$

Thus, if coordinate and momentum correlate with one another, then the product of their uncertainties increases by $K$ times compared to the value defined by the Heisenberg inequality.

Note, that due to the fact that coordinate and momentum do not commute, their quantum covariation can not be estimated by their sampling similar to the classical covariance. For calculating the corresponding estimate, one has to know a priori the state vector (wave function).

Let:

$$\psi(x) = \sqrt{\rho(x)} \exp(iS(x)), \tag{2.46}$$

where real functions $\rho(x)$ and $S(x)$ are correspondingly density and phase of the psi-function. Note that phase $S(x)$ is analogous to the classical action of a mechanical system.

Using the functions for density and phase it is simple to derive the following representation for the covariance of coordinate and momentum:

$$\text{cov}(x, p) = \frac{1}{2} \langle \psi | xp + px | \psi \rangle = \int x \frac{\partial S(x)}{\partial x} \rho(x) dx \tag{2.47}$$

The clearness of the obtained result is due to the fact that in classical mechanics the derivative of the action $\frac{\partial S}{\partial x}$ is the momentum.

## 2.5 Multidimensional uncertainty relation

Let us consider space of dimension $s$.

Let $\hat{x}_j, \hat{p}_j, \quad j = 1, ..., s$ be the corresponding operators of coordinate and momentum. The derivation of the uncertainty relation in the multidimensional case is similar to the one-dimensional case, but instead of the real number $\xi$ one has to introduce a real symmetric matrix $\Xi$ with elements $\xi_{j\sigma}, \quad j, \sigma = 1, ..., s$. The reason for this is to provide a geometrically invariant form in Hilbert space to the considered values. Actually, for a scalar $\xi$, the value like $(i\xi \hat{p}_\rho + \hat{x}_l)|\psi\rangle$ is not invariant, because the indices $\rho$ and $l$ are generally different. At the same time, for matrix $\Xi$ the value $(i\xi_{l\rho} \hat{p}_\rho + \hat{x}_l)|\psi\rangle$ is a ket-vector in Hilbert space (summation by the repeated index $\rho$ is assumed). Also let us introduce a real vector $\eta$ ($\eta_j \quad j = 1, ..., s$). Using this vector, one can transform the obtained ket-vector into a scalar by taking the inner product: $(i\xi_{l\rho} \hat{p}_\rho + \hat{x}_l)\eta_l|\psi\rangle$.

Now, let us consider the following wittingly non-negative expression (as usual, summation by repeating indices is assumed):

$$F(\xi) = \langle \psi | \eta_j \left( -i\xi_{j\sigma} \hat{p}_\sigma + \hat{x}_j \right) \left( i\xi_{l\rho} \hat{p}_\rho + \hat{x}_l \right) \eta_l | \psi \rangle \geq 0 \tag{2.48}$$

In expanded view we get:

$$F(\xi) = \langle \psi | \eta_j \eta_l \left( \xi_{j\sigma} \xi_{l\rho} \hat{p}_\sigma \hat{p}_\rho - i(\xi_{j\rho} \hat{p}_\rho \hat{x}_l - \xi_{l\rho} \hat{x}_j \hat{p}_\rho) + \hat{x}_j \hat{x}_l \right) | \psi \rangle \geq 0 \tag{2.49}$$

In order to use the fundamental commutation relations between the coordinate and momentum, we shall re-write the latter expression by substituting $j$ with $l$, and summing the obtained expression with the initial one. As observable variables we shall use centered coordinates and momentums (with zero mean values). As a result the condition will be obtained, according to which the following matrix expression is non-negatively defined:

$$\Xi \Sigma_p \Xi - \Xi + \Sigma_x \geq 0 \tag{2.50}$$

Remember that matrix $A$ with elements $a_{jk}$ is called non-negatively defined if for any vector $|z\rangle$:

$$\langle z | A | z \rangle = a_{jk} z_j^* z_k \geq 0 \tag{2.51}$$

In this inequality we have introduced coordinate and momentum covariation matrices. The elements of these matrices are defined by the following expressions:

$$(\Sigma_x)_{jl} = \langle \psi | \hat{x}_j \hat{x}_l | \psi \rangle \tag{2.52}$$

$$(\Sigma_p)_{jl} = \langle \psi | \hat{p}_j \hat{p}_l | \psi \rangle \tag{2.53}$$

Let us account for the fact that non-negative definiteness of a matrix allows one to get the square root of it. Remember that an arbitrary Hermitian matrix $A$ can be transferred to a diagonal form, i.e. it can be re-written as follows:

$$A = UDU^+, \tag{2.54}$$

where $U$ is a unitary matrix and $D$ is a real diagonal matrix.

In addition, if matrix $A$ is non-negatively defined, it has non-negative eigenvalues that form the diagonal of matrix $D$. In this case, the operation of matrix square root evaluation is well defined:

$$A^{1/2} = UD^{1/2}U^+ \tag{2.55}$$

Using the notion of a matrix square root the expression obtained above can be re-written in the form:

$$\left( \Xi \Sigma_p^{1/2} - \frac{1}{2}\Sigma_p^{-1/2} \right)\left( \Sigma_p^{1/2}\Xi - \frac{1}{2}\Sigma_p^{-1/2} \right) - \frac{1}{4}\Sigma_p^{-1} + \Sigma_x \geq 0 \tag{2.56}$$

The first summand is wittingly non-negatively defined (and it is equal to zero for $\Xi = \frac{1}{2}\Sigma_p^{-1}$). This implies that the expression $-\frac{1}{4}\Sigma_p^{-1} + \Sigma_x$ is non-negatively defined, i.e.

$$\Sigma_x \geq \frac{1}{4\Sigma_p} \tag{2.57}$$

The inequality derived is the desired multi-dimensional uncertainty relation. It has the following meaning: for any quantum state, the matrix equal to the difference $\Sigma_x - \frac{1}{4}\Sigma_p^{-1}$ between coordinate covariation matrix and one fourth of the matrix inverse to the momentum covariation matrix, is always non-negatively defined.

From these calculations it immediately follows that the inequality transforms into an equality if and only if the state vector satisfies the following condition for $\Xi = \frac{1}{2}\Sigma_p^{-1}$:

$$\left( i\xi_{l\rho}\hat{p}_\rho + \hat{x}_l \right)|\psi\rangle = 0 \tag{2.58}$$

Therefore, the corresponding state is Gauss-type with the covariance matrix $\Sigma_p$ in momentum representation and matrix $\Sigma_x = \frac{1}{4\Sigma_p}$ in coordinate representation.

We have limited ourselves to considering multi-dimensional uncertainty relation, which is a direct generalization of the one-dimensional Heisenberg's uncertainty relation. Other examples of generalized uncertainty relations, and, in particular, those connected with the Schrödinger-Robertson relation generalization, can be found in [19,20]

### 2.6. Fisher Information.

Let us consider a system that has a real psi-function: $\psi(x) = \sqrt{P(x)}$. For this system the mean momentum is equal to zero and the square of momentum is:

$$M(\hat{p}^2) = \int \frac{\partial}{\partial x}\sqrt{P(x)} \frac{\partial}{\partial x}\sqrt{P(x)}dx = \frac{1}{4}\int \frac{(P'(x))^2}{P(x)}dx \tag{2.59}$$

Here the apostrophe implies the derivation by $x$.

Let us introduce Fisher information related to the momentum variance:

$$I_x = 4D_p = 4M(\hat{p}^2) = \int \frac{(P'(x))^2}{P(x)} dx \tag{2.60}$$

Then the uncertainty relation can be expressed as the following relation: $D_x I_x \geq 1$.

The inequality derived is analogous to the Cramer-Rao inequality that is considered in the next section.

### 2.7. Cramer-Rao inequality

Let a quantum state depend on some real parameter $\theta$. We shall express it as:

$$\psi(x|\theta) = \sqrt{P(x|\theta)} \tag{2.61}$$

Let $\hat{\theta}$ be an unbiased estimator of an unknown parameter $\theta$, based on sampling $n$ in coordinate space i.e. $\hat{\theta} = \hat{\theta}(x_1,...,x_n)$. The condition for unbiasedness implies that the average value of estimate $\hat{\theta}$ is equal to the true value of parameter $\theta$, i.e.

$$M(\hat{\theta}) = \int P(x_1|\theta) \cdots P(x_n|\theta) \cdot \hat{\theta}(x_1,...,x_n) dx_1 \cdots dx_n = \theta \tag{2.62}$$

Some examples of unbiased estimators are well known estimates of mean value (expectation) and variance [14]:

$$\bar{x} = \frac{x_1 + ... + x_n}{n} \tag{2.63}$$

$$s^2 = \frac{1}{n-1} \sum_{k=1}^{n} (x_k - \bar{x})^2 \tag{2.64}$$

Let $p_\theta = -i \frac{\partial}{\partial \theta}$ be an operator that is canonically conjugate to parameter $\theta$.

Our next task is to evaluate the following relation called Cramer-Rao inequality:

$$D_\theta I_\theta \geq 1 \tag{2.65}$$

Here we have defined Fisher's information that has the form:

$$I_\theta = n \int \frac{(\partial P(x|\theta)/\partial \theta)^2}{P(x|\theta)} dx = n \int \left( \frac{\partial \ln P(x|\theta)}{\partial \theta} \right)^2 P(x|\theta) dx \tag{2.66}$$

Let us account for the fact that the sample size state vector can be defined as follows:

$$\psi(x_1,...,x_n) = \sqrt{P(x_1|\theta) \cdots P(x_n|\theta)} \tag{2.67}$$

Let us make expanded calculations. Let $\xi \frac{\partial \psi}{\partial \theta} + (\theta - \hat{\theta})\psi$ be a ket-vector, where $\xi$ is as earlier an arbitrary real parameter and $\xi \frac{\partial \psi^*}{\partial \theta} + (\theta - \hat{\theta})\psi^*$ is the corresponding bra-vector. The wittingly non-negative expression is:

$$F(\xi) = \int \left( \xi \frac{\partial \psi^*}{\partial \theta} + (\theta - \hat{\theta})\psi^* \right) \left( \xi \frac{\partial \psi}{\partial \theta} + (\theta - \hat{\theta})\psi \right) dx \tag{2.68}$$

Here to shorten the expression we assume that $dx = dx_1 \cdots dx_n$, $\psi = \psi(x_1,...,x_n)$
In expanded representation we have:

$$F(\xi) = a\xi^2 + b\xi + c \geq 0 \quad (2.69)$$

where

$$a = \frac{I_\theta}{4} = \int \frac{\partial \psi^*}{\partial \theta} \frac{\partial \psi}{\partial \theta} dx \quad (2.70)$$

$$b = \int (\theta - \hat{\theta}) \psi^* \frac{\partial \psi}{\partial \theta} + (\theta - \hat{\theta}) \frac{\partial \psi^*}{\partial \theta} \psi \, dx \quad (2.71)$$

$$c = \int (\theta - \hat{\theta})^2 \psi^* \psi \, dx = D_\theta \quad (2.72)$$

It can be demonstrated that $b = -1$. For this one has to use the expression for the derivative of the product as:

$$(\theta - \hat{\theta})\psi^* \frac{\partial \psi}{\partial \theta} + (\theta - \hat{\theta}) \frac{\partial \psi^*}{\partial \theta} \psi = \frac{\partial\left((\theta - \hat{\theta})\psi^* \psi\right)}{\partial \theta} - \psi^* \psi \frac{\partial}{\partial \theta}(\theta - \hat{\theta}) =$$

$$\frac{\partial\left((\theta - \hat{\theta})\psi^* \psi\right)}{\partial \theta} - \psi^* \psi \quad (2.73)$$

The integral of the first summand is equal to zero due to the unbiasedness of the estimator. As a result, accounting for the normalization condition, we get $b = -1$.

Finally, the Cramer-Rao inequality is obtained [14, 15, 22, 23]:

$$D_\theta \geq \frac{1}{I_\theta} \quad (2.74)$$

Note that we have made calculations not only for the case of real vector assumed in the classical mathematical statistics, but also for a more general case of complex psi-functions.

In this case Fisher information is equal to:

$$I_\theta = 4 \int \frac{\partial \psi^*}{\partial \theta} \frac{\partial \psi}{\partial \theta} dx \quad (2.75)$$

Fisher information is analogous to momentum variance and it differs from the latter only in multiplier 4 and the fact that the integrand value is differentiated with respect to the parameter, instead of the coordinate.

For the case of real psi-functions, it is easy to show that the expression for Fisher information presented above (2.66) takes place. In the derivation of the result, one has to apply the property of Fisher information additivity (the information from $n$ independent representatives is $n$ times greater than the information from one representative).

The derived inequality is clearly the strictest for the case when Fisher information $I_\theta$ is minimal. Similar to the case of Heisenberg's uncertainty relation it can be shown that adding an arbitrary phase multiplier to the real psi-function can not reduce the Fisher information.

It has been demonstrated above that the uncertainty inequality transforms into an equality for the case of a Gauss-type state. An analogous result is true for the case of Cramer-Rao inequality. It transforms into an equality only for estimates that have normal distribution. Such estimates are called effective estimates.

The logic presented above allows one to easily derive the Cramer-Rao inequality for biased estimators also. Then it has the following form:

$$M(\theta - \hat{\theta})^2 \geq \frac{(1 + \beta'(\theta))^2}{I_\theta} \quad (2.76)$$

where $\beta(\theta) = M(\hat{\theta}) - \theta$ \quad (2.77)

is the bias of an estimator.

Note that in the inequality presented, the left part corresponds to the dispersion of the sample estimator $\hat{\theta}$ in respect to the real value $\theta$ instead of the ordinary variance.

## 2.8. Multi-dimensional Cramer-Rao inequality and the root estimator

*«Get at the root of it!» (Koz'ma Prutkov «Thoughts and Aphorisms», №228).*

The Cramer-Rao inequality, as well as the uncertainty inequality, can be generalized to the multi-dimensional case.

It can be demonstrated that for any unbiased estimator $\hat{\theta}$ of an unknown multi-dimensional parameter $\theta$ the matrix $\Sigma_\theta - I_\theta^{-1}$ is non-negatively defined:

$$\Sigma_\theta - I_\theta^{-1} \geq 0 \qquad (2.78)$$

In case of the estimators that are closed to effective estimators, the corresponding difference is equal approximately to zero. An example of such estimators is the maximum likelihood estimators that are asymptotically effective [15, 22, 23].

Here $\Sigma_\theta$ - is the covariance matrix of estimator $\hat{\theta}$. The elements of Fisher information matrix $I_\theta$ can be represented as:

$$(I_\theta)_{jk} = n \int \frac{\partial \ln P(x|\theta)}{\partial \theta_j} \frac{\partial \ln P(x|\theta)}{\partial \theta_k} P(x|\theta) dx \qquad (2.79)$$

From the point of view of quantum informatics, it is very important that the expression for the Fisher information matrix is simplified a lot if the psi-function is introduced (for the sake of simplicity we assume that it is real-valued) [6].

$$(I_\theta)_{jk} = n \int \frac{\partial \psi(x|\theta)}{\partial \theta_j} \frac{\partial \psi(x|\theta)}{\partial \theta_k} dx \qquad (2.80)$$

For statistical problems the matrix inverse to the Fisher information matrix is of fundamental importance.

Due to the difficultness of expression (2.79) for a multi-dimensional Fisher information matrix the estimators of the inverse matrix derived on its basis are usually ill- defined. The only exception from the rule is the so-called root estimator that is based on the introduction of a psi-function.

Let us briefly present the results derived. A more complete analysis can be found in [6, 7].

Let a decomposition of a psi-function on a orthonormal base function set $\varphi_j(x)$ $j = 0,1,..., s-1$ be of the form:

$$\psi(x) = \sqrt{1 - (c_1^2 + ... + c_{s-1}^2)} \varphi_0(x) + c_1 \varphi_1(x) + ... + c_{s-1} \varphi_{s-1}(x) \qquad (2.81)$$

Here we have excluded the coefficient $c_0 = \sqrt{1 - (c_1^2 + ... + c_{s-1}^2)}$ out of the estimated parameters, because due to the normalization condition it can be expressed by other coefficients.

The values $c_1, c_2, ..., c_{s-1}$ are independent estimated parameters.

In case of the root decomposition (2.81) the information matrix $I_{ij}$ has the size $(s-1) \times (s-1)$ and it can be expressed in the following simple form:

$$I_{ij} = 4n \left( \delta_{ij} + \frac{c_i c_j}{c_0^2} \right), \text{ where } c_0 = \sqrt{1 - (c_1^2 + ... + c_{s-1}^2)} \qquad (2.82)$$

One remarkable feature of the derived result is its independence on the choice of base functions. It appears that only the root density estimator has this property.

For estimators that are close to the effective ones, the covariation matrix of the state vector estimator is approximately a matrix that is inverse to the Fisher information matrix.

$$\Sigma(\hat{c}) = I^{-1}(c) \qquad (2.83)$$

The components of this matrix are:

$$\Sigma_{ij} = \frac{1}{4n}(\delta_{ij} - c_i c_j) \quad i,j = 1,\ldots,s-1 \qquad (2.84)$$

The derived matrix can be expanded by adding the covariances of components $c_0$ of the state vector with other components of the vector to its covariances. It appears that the aggregate covariance matrix has the same form as (2.84), where $i,j = 0,1,\ldots,s-1$.

Thus, the statistical model based on a psi-function and root decomposition introduction and the methods of quantum informatics is in some way preferred to all other feasible models. The advantages of this model are the simplicity of its notations, the general nature of its results and the good-definiteness of its computational properties. According to Dirac "The Nature could not have ignored such a beautiful mathematical model".

## 3. The postulates of quantum informatics

*"Every tailor has his own views on art!"* (Koz'ma Prutkov «Thoughts and Aphorisms», №151).

The postulates of quantum informatics should make apparent all of the most profound and the most fundamental ideas of quantum theory. There are different opinions on the issue of which notions of quantum theory should be considered the main ones. In this case, it is quite interesting to explore the evolution of Dirac's opinions on the paradigm of quantum physics. The actual Dirac in 1930 in his outstanding work "The principles of quantum mechanics" [24] according to Von Neumann "provided us with such a brief and an elegant description of quantum mechanics that is unlikely to be ever overcame" (a translation from the Russian edition [25], p.10).

Note that many other outstanding scientists shared similar delights of Dirac's wordings of the principal notions of quantum theory. That makes even more valuable what Dirac writes himself in 1972 in his work "Relativity theory and quantum mechanics" [26] about the evolution of his own views on the issue.

"A question of whether the non-commutativity is the principal notion of quantum mechanics arises. Previously, I had believed this to be true, but recently I started to doubt this point. May be there is some other more profound notion than the notion of non-commutativity, some deeper modification of our understanding of the world that quantum mechanics brings us" (a translation from the Russian edition [26], p.148).

Note that the notion of non-commutativity was very appealing to Dirac. This very notion allowed Dirac to formulate the notion of Poisson quantum brackets instead of the classical ones, which, in its turn, allowed him to beautifully and elegantly transform classical mechanics into quantum mechanics. And then, forty years after his pioneering works, Dirac concludes that there must be a deeper idea, which is the idea of probability amplitudes existence. The following words are marked in courier by Dirac: *"I believe that the notion of probability amplitudes is the most fundamental notion of quantum theory"* (a translation from the Russian edition [26], p.148).

This framework exactly corresponds to the spirit of the sixth Hilbert's problem – a construction of a special type of probability theory based on geometry.
It is worth considering what the "Principles of quantum mechanics" would have been like if the young Dirac shared his own ideas of the more senior age. It appears that it is not obligatory to consider the procedure of Dirac's canonical quantization that is based on Poisson quantum brackets to transform classical mechanics into quantum mechanics. It is sufficient to stand to the concept of probability amplitudes and the statistical requirement of the coincidence, on the average, of the results of the new and the old theories. (for more detail see [8, 9, 10]).

Basing on the description provided above, let us formulate the first postulate.

**Postulate 1.** *The principal object of quantum informatics is a quantum system. The evolution of the quantum system is described by probability amplitudes. Probability amplitudes construct a state vector in Hilbert space.*

Hilbert space is a linear vector space. The property of linearity implies the principle of superposition validity. Therefore, if $|a\rangle$ and $|b\rangle$ are the vectors that describe some states of a system, then their arbitrary linear superposition $c_1|a\rangle + c_2|b\rangle$ (where $c_1, c_2$ are arbitrary complex numbers) is also a feasible state of the system (the superposition principle).

State vector as a geometric object in Hilbert space can be defined in different equivalent representations that are unitary interconnected similar to the case of when the description of the behaviour of two classical objects in

Euclidian space can be studied in different coordinate representations that are interconnected by orthogonal transformations. This reasoning lies in the basis of the following postulate.

**Postulate 2.** *Probability amplitudes as state vector coordinates in Hilbert space can be defined in different equivalent representations. The equivalent representations are interconnected by unitary transformations. A unitary transformation in time describes the evolution of the quantum system.*

A unitary transformation can be written symbolically by the following matrix equation:

$$|\psi'\rangle = U|\psi\rangle \quad (3.1)$$

Any unitary matrix $U$ can be represented as:

$$U = \exp(iH), \quad (3.2)$$

where $H$ is a Hermitian matrix.

Due to the time homogeneity, the unitary transformation in time has to satisfy the following relation:

$$U(t_1 + t_2) = U(t_1)U(t_2) \quad (3.3)$$

The matrix exponent that satisfies the condition of time homogeneity has to be of the form:

$$U = \exp(iHt) \quad (3.4)$$

The Hermitian operator $H$ introduced in the way is called the Hamiltonian.

The latter equation implies that the unitary evolution of quantum states has to be defined by Schrödinger equation:

$$i\frac{\partial \psi}{\partial t} = H\psi \quad (3.5)$$

A state vector is an objective statistical characteristic of a quantum system and it has to allow for the possibility of a statistical study. However, for such an exploration a number of statistical ensemble representatives are needed instead of only one representative. In the ensemble every representative has to be prepared equally and, thus, all representatives should be in the same quantum state.

It is not sufficient to make measurements in one basis. One has to perform the measurements in various unitary-interconnected bases. The results of such measurements are in accordance with the following postulate.

**Postulate 3.** *Measurements conducted in different unitary interconnected basis representations generate a set of mutually-complementary statistical distributions. For a fixed basis the square of absolute value of probability amplitude defines the probability of quantum system detection in a corresponding basis state.*

Postulates 2 and 3 are closely related with one another. From one side, postulate 3 serves for "materializing" the results of transformations that are defined by postulate 2 (that is evident). From the other side, if one is to perform measurements in accordance with postulate 3, one has to ensure that such measurements yield the most complete description of the events. This can not be assured if we use only one of the representations. Therefore, to conduct measurements in accordance with postulate 3, one has to apply postulate 2 as well, performing the transfer between two representations (that is less evident).

For every representative of a statistical quantum ensemble, a choice is to be made: whether to conduct the measurement in the initial representation or whether to switch to the other representation and perform the measurement only then.

Only the aggregate of measurements in various mutually-complementary representations can provide a complete picture of a quantum state in the experimental framework.

By the reasoning provided above, we assume that a representative that is once measured will not be measured any more. Even if such a measurement was to be performed it would yield the information not on the initial quantum state but on the state that resulted after the first measurement. That is the property of quantum state reduction. "No matter how hard one tries, but an egg can not be incubated twice" (Koz'ma Prutkov "Thoughts and aphorisms", №258). A modern variant of the aphorism could sound like: "No matter how hard one tries, but a quantum bit can not be «*inqubited*» twice" ☺.

Considering quantum states of composite systems we essentially come to the notion of tensor product of sub-system states. For instance, let us consider a system of two two-level quantum systems (qubits). It is natural to assume that the system should contain the following four base states as the feasible states.

$|00\rangle$ - both qubits are in state $|0\rangle$,

$|01\rangle$ - the first qubit is in state $|0\rangle$, the second one – in state $|1\rangle$,

$|10\rangle$ - the first qubit is in state $|1\rangle$, the second one – in state $|0\rangle$,

$|11\rangle$ - both qubits are in state $|1\rangle$.

The four base vectors presented produce a Hilbert space of dimension 4. It implies that a system of two qubits can be not only in one of the states, but also in any superposition of the states.

$$|\psi\rangle = c_{00}|00\rangle + c_{01}|01\rangle + c_{10}|10\rangle + c_{11}|11\rangle \qquad (3.6)$$

Thus, the following postulate is quite natural:

**Postulate 4** *The space of a composite system state is produced by the tensor product of the states of individual systems.*

For instance, $n$ qubits that are considered as a whole produce $2^n$ base states and consequently a Hilbert space of dimension $2^n$. An arbitrary state vector in the space is defined by $2^n$ complex probability amplitudes. Note that for independent states of every qubit, there are only $2n$ complex probability amplitudes, which makes a much smaller amount for large $n$. The difference $2^n - 2n$ is due to the specific quantum resource, namely entanglement. A quantum state is called to be entangled if it can not be presented by individual sub-systems states. The entanglement is the notion that is to make quantum computers exponentially more powerful compared to their classical counterparts.

We shall note that Postulate 4 makes unavoidable a probabilistic realization of a quantum informational model. For example, for $n = 1000$ qubits there is a state that is described by $2^{1000} \approx 1{,}07 \cdot 10^{301}$ complex numbers. For the whole Universe that has "only" $\sim 10^{78}$ nucleons at its disposable there is no way to record the state on any material object.

**Discussion**

Let us briefly discuss the history of the sixth Hilbert's problem evolution in the 20[th] century.

First of all, basing on his thesis of the necessity of combining probability theory research with kinetic gas theory, Hilbert applied his theory of integral equations to Boltzmann's kinetic equation. This allowed Hilbert to discover an effective way of the kinetic equation approximate solution [27].

The Boltzmann's kinetic equation was an example to Hilbert of an equation that was integral-type by its point in the sense that it could not have been transferred to any type of differential equations.

The emergence of quantum mechanics with the works by Heisenberg [28], Born and Jordan [29] and Heisenberg, Born and Jordan [30] in 1925, stimulated Hilbert to explore the mathematical tools of the new theory. He worked on the problem together with his assistants – von Neumann and Nordhame. The results of the study were published in [31].

At the same time, the cooperation with Hilbert stimulated von Neumann to perform systematical research on the mathematical description of quantum theory. After a few years' time, a book [25] was published that summarized the main results if the work. Ever since, this book has been the principal work on this issue related to the mathematical aspects of quantum mechanics.

In his monograph von Neumann extended the concept of Hilbert space as a place where quantum events take place. He also introduced the notion of density matrix, developed the theory of quantum measurements, based on the decomposition of unity and conducted a research on quantum statistical mechanics grounding.

Von Neumann attempted to express his vision of the fundamental statistical bases of quantum mechanics in his famous theorem that states the impossibility of introducing hidden parameters to the structure of quantum theory.

According to von Neumann, this theorem should have separated quantum and classical statistical theories. This theorem had not been argued for a long time, until it was heavily criticized by Bell [32]. The research conducted by Bell, resulted in a set of inequalities that were called after his name. These inequalities demonstrate the impossibility to explain the results of statistical experiments on quantum objects by the means of classical probabilistic space. In this framework, Bell's inequalities express quantitatively what was formulated by von Neumann qualitatively. However, we believe that the theorem of the impossibility of introducing hidden parameters to quantum mechanics, as well as Bell's inequalities, despite all their importance, do not play that fundamental role that is often believed to be the case. Indeed, when many authors stress the Bells inequalities and the hidden parameter issue they leave out more fundamental results of quantum theory, related to Bohr's complementarity principle and Heisenberg's uncertainty relations. Actually, the complementarity principle that is based on rejecting the axiom of composite random values of classical probability theory implies that there is a more fundamental notion – the state vector of a quantum system. The concept of mutually-complementary distributions leads to the fundamental uncertainty relation. In particular, Heisenberg's inequality reveals that the product of variances of coordinate and momentum distributions can not be lower than the value determined by Planck's constant. In classical statistics, where all observable random values can be defined simultaneously, such an effect can not take place in principle.

The mathematical tools that were developed by von Neumann were significantly modified and generalized by other authors. For example, in the contemporary theory of quantum measurements general decompositions of the unity are considered, namely Positive Operator- Valued Measure – POVM (for more details see [20, 33, 34])

The history of Kolmogorov axiomatic is described in [35].

## Conclusion:

Let us formulate the main results of the work.
1. It was demonstrated that quantum informatics is a natural solution of the 6-th Hilbert's problem that is related to constructing axiomatic probability theory describing micro-world.
2. Basing on the Fourier transform theory, in accordance with the Bohr's principle of complementarity mutual coordinate and momentum distributions were studied. It was shown that the characteristic function may be considered as a convolution of the psi-function. The reasoning for momentum and coordinate operators introduction is provided.
3. Hilbert space geometry is presented in probabilistic framework. It was demonstrated that Cauchy-Bunyakowsky inequalities, uncertainty relations (one-dimensional and multi-dimensional) and Cramer-Rao inequalities can considered from the same quantum-informational standpoints.
4. The main postulates of quantum informatics as a natural science were formulated. It was shown that quantum mechanics is a root statistical model (based not on the actual probabilities, but on the "square roots" of them).

The author is grateful to K.A. Valiev, Yu.I. Ozhigov and A.V. Shevrev for valuable discussions concerning the problem.

## REFERENCES:

1. *Nielsen M.A., Chuang I.L.* Quantum Computation and Quantum Information // Cambridge University Press. 2001.
2. *Valiev K.A., Kokin A.A.* Quantum Computers: hopes and reality (in Russian). Second edition. Moscow. RCD. 2004. 320 p.
3. The Physics of Quantum Information. Quantum Cryptography. Quantum Teleportation. Quantum Computation // Bouwmeester D., Ekert A, Zeilinger A. (Eds.). Springer. 2001.
4. *Cohen- Tannoudji C., Diu B., Laloö F.* Quantum Mechanics. Vol. I, II. John Wiley and Sons. New York. 1977.
5. *Landau L. D., Lifschitz E. M.* Quantum Mechanics (Non-Relativistic Theory). 3rd ed. Pergamon Press. Oxford. 1991.
6. *Bogdanov Yu. I.* Fundamental problem of statistical data analysis: Root approach. M.: MIEE, 2002. 84 p.; *Bogdanov Yu. I.* Statistical inverse problem: Root Approach // LANL E-print, 2003, arXiv: quant-ph/0312042. 17 p.; LANL E-print, 2002, arXiv: physics/0211109. 39p.
7. *Bogdanov Yu.I.* Quantum mechanical view of mathematical statistics // Proceedings of SPIE. 2006. V.6264. 62640E; LANL E-print, 2003, arXiv: quant-ph/0303013. 26 p.
8. *Bogdanov Yu.I.* Root estimator of quantum states // LANL E-print, 2003, arXiv: quant-ph/0303014. 26 p.
9. *Bogdanov Yu. I.* Fundamental Notions of Classical and Quantum Statistics: A Root Approach // Optics and spectroscopy. 2004. V.96. №5. P. 668-678.
10. *Bogdanov Yu. I.* Quantum States Estimation: Root Approach // Proceedings of SPIE. 2004. V.5401. P.741-750. ; LANL E-print, 2003, arXiv: quant-ph/0310011. 10 p.
11. *Hilbert D.* Mathematische Probleme // Nachr. Ges. Wiss. Göttingen 1900, s.253-297. (Translation in to Russian: *Hilbert D.* Mathematical problems- Selected Scientific Papers. Moscow. Factorial. 1998. V.2. P.401-436)
12. *Zee H.D.* Roots and Fruits of Decoherence // Seminaire Poincare. 2005. p.115-129.; *Zurek W.H.* Decoherence and the Transition from Quantum to Classical- Revisited// Ibid. p.1-23.
13. *Boltzmann L.* Further research of heat equilibrium between gas molecules // Selected Scientific Papers. Moscow. Nauka. 1984. (in Russian)
14. *Cramer H.* Mathematical Methods of Statistics. Princeton University Press. Princeton. 1946. (Russian Edition. Moscow. Mir. 1975)
15. *Kendall M., Stuart A.* The Advanced Theory of Statistics. Inference and Relationship. U.K. Charles Griffin. London. 1979.
16. *Gnedenko B.V.* The Theory of Probability. Chelsea. New York. 1962.
17. *Lukacs E.* Characteristic Functions. London. Charles Griffin & Company Limited. 1960.216 p.

18. *Prokhorov L.V.* Quantum Mechanics – Problems and Paradoxes. Saint Petersburg. 2003. (in Russian)
19. *Robertson H.P.* An Indeterminacy Relation for Several Observables and Its Classical Interpretation // Phys.Rev. 1934. V.46. P.794-801
20. *Holevo A.S.* Statistical Structure of Quantum Theory. Berlin. Springer- Verlag. 2001. (Russian edition- RCD. 2003)
21. *Bogdanov A.Yu., Bogdanov Yu.I., Valiev K.A.* Schmidt information and entanglement in quantum systems// Moscow State University Proceedings. Ser.15. Computational Mathematics and Cybernetics. 2006 (accepted for publication); quant-ph/0512062 *Bogdanov A.Yu., Bogdanov Yu.I., Valiev K.A.* Schmidt Modes and Entanglement in Continuous-Variable Quantum Systems // Russian Microelectronics. 2006. №1.; quant-ph/0507031.
22. Probability theory and mathematical statistics. Encyclopedia. // Prokhorov Yu.V. (Ed.). The Large Russian Encyclopedia. 1999. 911p. (in Russian)
23. *Kryanev A.V., Lukin G.V.* Mathematical methods of uncertain data processing. Moscow, Fizmatlit. 2003. 216 p. (in Russian)
24. *Dirac P.A.M.* The principles of quantum mechanics. 4ed., Oxford, 1958 (Translation in to Russian: *Dirac P.A.M.* Collected Scientific Papers. Moscow. Fizmatlit. 2002. V.1. P.7-320)
25. *Von Neumann J.* Mathematische Grudlagen der Quantenmechanik. Berlin. Springer. 1932. (Russian edition- Moscow. Nauka. 1964)
26. *P.A.M. Dirac* Relativity and Quantum Mechanics // Fields and Quanta. 1972. V3. P.139-164. (Translation in to Russian: *Dirac P.A.M.* Collected Scientific Papers. Moscow. Fizmatlit. 2004. V.3. P.141-152)
27. *Hilbert D.* Grundzüge einer allgemeinen Theorie der linearen Integralgleichungen (Basics of the general theory of linear integral equations)// Leipzig und Berlin: Verlag und Druck von B.G. Teubner. 1924. (Translation in to Russian: *Hilbert D.* Selected Scientific Papers. Moscow. Factorial. 1998. V.2. P.350-364)
28. *Heisenberg W.* Über quantentheoretische Umdeutung kinematischer und mechanischer Beziehungen // Zs. Phys., 1925, **33**, 879-893. (Translation in to Russian: *Heisenberg W.* Selected Scientific Papers. Moscow. URSS. 2001. P.86-98)
29. *Born M, Jordan P.* Zur Quantenmechanik // Zs. Phys., 1925, **34**, H.I-II, 858-888. (Translation in to Russian: *Heisenberg W.* Selected Scientific Papers. Moscow. URSS. 2001. P.99-126)
30. *Heisenberg W, Born M, Jordan P.* Zur Quantenmechanik II // Zs. Phys., 1926, **35**, 8-9, 557-615. (Translation in to Russian: *Heisenberg W.* Selected Scientific Papers. Moscow. URSS. 2001. P.127-175)
31. *Hilbert D., Neumann J., Nordheim L.* Über die Grundlagen der Quantenmechanik. 1928. Math. Ann. Bd. 98. s. 1-30.
32. *Bell J.S.* Speakable and unspeakable in quantum mechanics. Collected papers on quantum philosophy. Cambridge University Press. 1993.
33. *Holevo A.S.* Introduction to quantum information theory. MTsNMO. Moscow. 2002 (in Russian)
34. *Holevo A.S.* Probabilistic and Statistical Aspects of Quantum Theory (Russian Edition. RCD. 2003)
35. *Gnedenko B.V.* On the sixth Hilbert's problem// Hilbert problems. A collection of articles edited by Aleksandrov P.S. Moscow. URSS. 2000. P.117-119 (in Russian)